\documentclass[aps,prl,twocolumn,groupedaddress,showpacs]{revtex4}
\usepackage{epsfig}

\begin{document}


\title{Atomic scale elastic textures coupled to electrons in superconductors}

\author{K. H. Ahn, Jian-Xin Zhu, Z. Nussinov, T. Lookman, A. Saxena, A. V. Balatsky, A. R. Bishop}
\affiliation{Theoretical Division, Los Alamos National
Laboratory, Los Alamos, New Mexico 87545}
\date{\today}

\begin{abstract}
We present an atomic scale theory of lattice distortions using
strain related variables and their constraint equations. Our
approach connects constrained {\it atomic length} scale variations
to {\it continuum} elasticity and describes elasticity at all
length scales. We apply the general approach to a two-dimensional
square lattice with a monatomic basis, and find the atomic scale
elastic textures around a structural domain wall and a single
defect, as exemplary textures. We clarify the microscopic origin
of gradient terms, some of which are included phenomenologically
in Landau-Ginzburg theory. The obtained elastic textures are used
to investigate the effects of elasticity-driven lattice
deformation on the nanoscale electronic structure in
superconductor by solving the Bogliubov-de Gennes equations with
the electronic degrees of freedom coupled to the lattice ones. It
is shown that the order parameter is depressed in the regions
where the lattice deformation takes place. The calculated local
density of states suggests the electronic structure is strongly
modulated as a response to the lattice deformation--- the
elasticity propagates the electronic response over long distances.
In particular, it is possible for the trapping of low-lying
quasiparticle states around the defects. These predictions could
be directly tested by STM experiments in superconducting
materials.

\end{abstract}


\maketitle

\newpage
In many complex electronic materials such as cuprates, manganites,
ferroelastic martensites, and titanates,  unexpected multiscale
modulations of charge, spin, polarization, and strain variables
have been revealed by high resolution microscopy~\cite{Multi1}. It
is increasingly evident that the nonuniform textures found in
these doped materials have intrinsic origins: they arise from
coupling between various degrees of freedom. The textures
fundamentally affect local and mesoscopic electronic, magnetic and
structural properties, which are central to the functionality of
correlated electronic materials. There is ample evidence for
significant coupling amongst the electronic degrees of freedom
with the lattice distortions in cuprates, manganites, and
ferroelectrics. The charge carrier doping can act as a local
stress to deform surrounding unit cells~\cite{Multi1}. We might
employ a Landau-Ginzburg (LG) theory to study the coupling between
the electronic (Cooper pair) and lattice (strain tensor) degrees
of freedom in superconductors. However, the LG theory can only
rigorously describe the long wavelength behavior. New generations
of experimental tools to probe individual atoms and local
environments~\cite{Stemmer95} and the aforementioned growing
interest in complex functional materials, emphasize the importance
of accurately describing the local electronic properties and
lattice distortion at the atomic scale. In this work, we first
present a microscopic description of elasticity.
We introduce appropriate {\em inter}-cell and {\em intra}-cell
distortion modes and show how the form of the elastic energy
recovers the correct phonon spectra. The discreteness of the
lattice, choice of modes and constraints among them give rise to
an anisotropic gradient expansion for the elastic energy. This
leads to interesting elastic domain wall and defect textures. We
then couple these textures to the electronic degrees of freedom
and study microscopically the influence of strain on electronic
wavefunctions in both $s$ and $d$-wave superconductors.

\begin{figure}
\leavevmode
\epsfxsize5.5cm\epsfbox{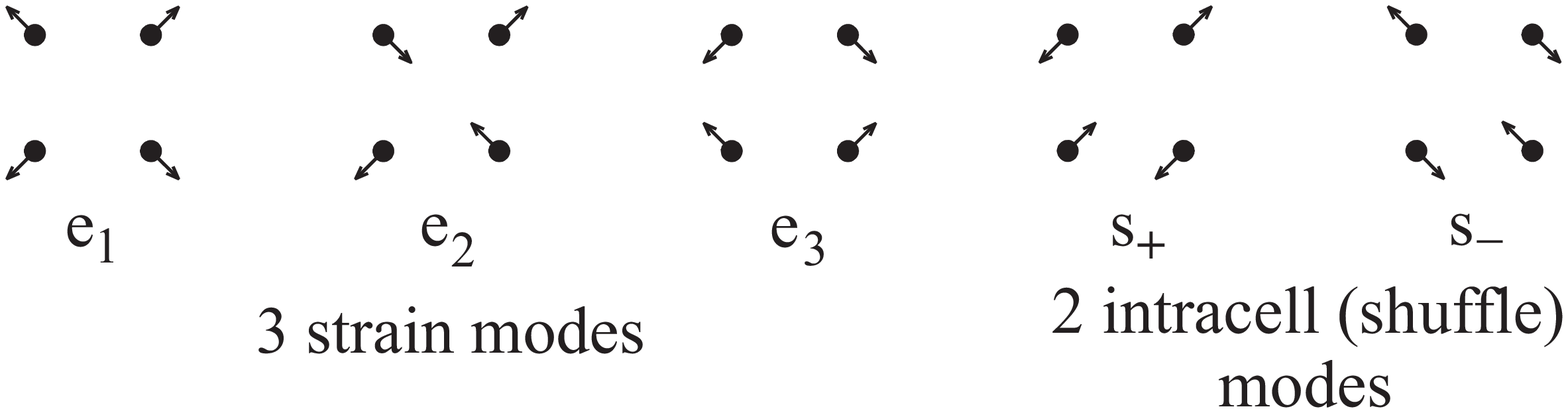}
\caption{\label{fig:squaremode} Normal distortion modes for
a square object of four atoms in
2D.
}
\end{figure}

Our approach is general, but we illustrate it here in detail for
the simplest case, namely a square lattice in two-dimensional (2D)
space with a monatomic basis. We find that the most convenient
strain-related variables for atomic scale distortions are the
normal distortion modes
 of an elementary square object of
four atoms (Fig.~\ref{fig:squaremode}).  The first three
distortion modes in Fig.~\ref{fig:squaremode} correspond to the
usual dilatation ($e_1$), shear ($e_2$), and deviatoric ($e_3$)
strains of the continuum elasticity theory for a square
lattice~\cite{Shenoy99}. The next two degenerate modes in
Fig.~\ref{fig:squaremode}, $s_+$ and $s_-$, correspond to the
``intracell'' or ``shuffle'' modes of the square
lattice~\cite{Barsch84}, which are  absent in continuum elasticity
theory. Our approach uses these five distortion variables defined
for each plaquette of four atoms at $\vec{i}$, $\vec{i}+(10)$,
$\vec{i}+(11)$, and $\vec{i}+(01)$, where $\vec{i}$ represents the
coordinate of the lattice points, to describe the elastic
energy~\cite{definition}.

Since the five variables
are derived from
two displacement variables for each lattice site,
they are related by three
constraint equations.
By representing $e_1$, $e_2$, $e_3$, $s_+$, and $s_-$
in terms of displacement variables $d^x$ and $d^y$ in $k$ (wavevector) space
and eliminating $d^x$ and $d^y$,
the constraint equations
are obtained.
One of them
is the {\it microscopic} elastic compatibility equation,
which relates strain modes:
\begin{eqnarray}
& &(1-\cos k_x \cos k_y) e_1 (\vec{k})
- \sin k_x \sin k_y e_2 (\vec{k})  \nonumber \\
& &+ (\cos k_x - \cos k_y) e_3 (\vec{k}) = 0. \label{eq:const.1}
\end{eqnarray}
The other two relate the intracell and the strain modes:
\begin{eqnarray}
&2 \cos \frac{k_x}{2} \cos \frac{k_y}{2} s_{\pm} (\vec{k}) \mp i
\sin \left(\frac{k_x \pm k_y}{2} \right)
e_1 (\vec{k}) & \nonumber \\
&\pm i \sin \left(\frac{k_x \mp k_y}{2} \right) e_3 (\vec{k}) = 0.
& \label{eq:const.2}
\end{eqnarray}
These constraints generate {\it anisotropic} interactions (from
the lattice symmetry) between atomic scale strain fields, similar
to the compatibility equations in continuum
theory~\cite{Shenoy99}, but now including the intracell modes. In
the long wavelength limit, our description naturally reproduces
the continuum results: For $\vec{k} \rightarrow 0$, the above
constraint equations
can be written in real space as
\begin{eqnarray}
&&\nabla^2 e_1(\vec{r})
-2 \nabla_x \nabla_y  e_2 (\vec{r})
+ (\nabla_y^2-\nabla_x^2) e_3(\vec{r}) = 0, \label{eq:comp} \\
&&s_{\pm}(\vec{r})= \frac{1}{4}\left[ \left( \nabla_y  \pm
\nabla_x \right) e_1(\vec{r}) + \left(  \nabla_y \mp \nabla_x
\right) e_3(\vec{r}) \right]. \label{eq:s.and.e}
\end{eqnarray}
Equation~(\ref{eq:comp}) is the usual compatibility equation in
continuum theory. Equation~(\ref{eq:s.and.e}) shows that {\it the
spatial variations of strains always generate intracell modes},
the magnitudes of which vanish as the inverse of the length scale
of the strain mode variations. In continuum LG theory, the energy
associated with the gradient of strains is responsible for domain
wall energies as, e.g., in structural phase
transitions~\cite{Barsch84}. The above result shows that the
intracell modes are the origin of such energy terms. Since our
strain-related variables become identical to conventional strain
variables for $k\rightarrow 0$ , various length scale lattice
distortions may be described in a {\it single} theoretical
framework. This makes it possible to study typical multiscale
situations where both short- and long-wavelength distortions are
important. It also provides a natural framework for incorporating
interactions between atomic scale strain-related fields coupled to
other degrees of freedom in functional materials (below).

The following analysis of the simple harmonic elastic energy for
the square lattice further exemplifies the utility of these
variables. We consider the simplest energy expression by
approximating the total elastic energy as the sum of the elastic
energy of each square:
\begin{equation}
E_{\text{sq.lat}}=\sum_{\vec{i}}\{ \sum_{n=1,2,3}
\frac{1}{2} A_n [e_n(\vec{i})]^2 + \sum_{m=+,-}
\frac{1}{2} B  [s_m(\vec{i})]^2 \}, \label{eq:E.sq}
\end{equation}
where $A_n$ and $B$ denote elastic moduli
and `intracell modulus', respectively.
Since some of the atomic pairs are shared by two
square plaquettes of atoms,
the parameters in
Eq.~(5) should be
appropriately renormalized.
A robust way to determine the parameters is
to compare the phonon spectrum of our model
with experimental data.

\begin{figure}
\leavevmode \epsfxsize6.5cm\epsfbox{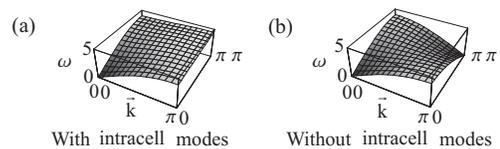}
\caption{\label{fig:phonon} An example of calculated phonon
spectra (a) with and (b) without intracell modes for a 2D square
lattice with a mono-atomic basis. The upper phonon branch is shown
for both cases. ($M=\hbar=1$). }
\end{figure}

The phonon spectrum has been obtained~\cite{Ahn02}. A typical
spectrum (upper branch) for $A_1$=5, $A_2$=4, $A_3$=3, and $B$=5
is shown in Fig.~\ref{fig:phonon}(a).
At $\vec{k}=(\pi,\pi)$, the distortion
is a pure intracell mode, and
the energy depends only on
the intracell mode modulus $B$.
Therefore, as shown in Fig.~\ref{fig:phonon}(b),
$\omega (\pi,\pi)$ vanishes
without the intracell mode ($B$=$0$),
which is unphysical.

We apply our formalism to obtain the domain wall solution for the
atomic displacements between two homogeneous strain states (a
``twin boundary'') due to a phase transition to a rectangular
lattice. We then compare the solution to that obtained from
continuum theory, where discreteness effects are neglected~\cite
{Barsch84}. With elastic energy
$E_{\text{rec}}=E_{\text{rec}}^{(1)}+E_{\text{rec}}^{(2)}$,
\begin{eqnarray}
E_{\text{rec}}^{(1)}&=&\sum_{\vec{i}}
\frac{1}{2} A_1 e_1(\vec{i})^2 +
\frac{1}{2} A_2 e_2(\vec{i})^2 +
\frac{1}{2} B [ s_+(\vec{i})^2 + s_-(\vec{i})^2 ],
\nonumber \\
E_{\text{rec}}^{(2)}&=&\sum_{\vec{i}} -\frac{1}{2} A'_3 e_3(\vec{i})^2 +
\frac{1}{4} F_3 e_3(\vec{i})^4,
\end{eqnarray}
the degenerate ground state of $E_{\text{rec}}$ is a uniform state
with $e_3$=$\pm \sqrt{A'_3/F_3}$, and $e_1$=$e_2$=$s_+$=$s_-$=0.
To study the domain wall between these two degenerate rectangular
ground states, we consider $e_3(\vec{i})$ as the order parameter
and minimize $E_{\text{rec}}^{(1)}$ with respect to the other
variables, using the constraint equations [Eqs. (1) and (2)] and
the method of Lagrange multipliers. We obtain
$E_{\text{rec,min}}^{(1)}=\sum_{\vec{k}} \frac{1}{2} e_3(-\vec{k})
U(\vec{k}) e_3(\vec{k})$, where $U(\vec{k})$ is given in
Ref.~\cite{Ahn02}.

With $k_x$=$k \cos \theta $ and $k_y$=$k \sin \theta$, the
expansion of $U(k,\theta)$ about $k$=0 yields
$U(k,\theta)=U_{0}(\theta) + U_{2}(\theta)k^{2} + O(k^4)$, where
$U_{0}(\theta) = A_1 A_2 \cos^2 2 \theta /(A_1 \sin^2 2\theta +
A_2)$, and $U_{2}(\theta)=\sin^2 2 \theta [ 6 A_1 A_2 B \sin^2 2
\theta + 4 A_1 A_2 ( A_1 + A_2) \cos^2 2 \theta + 3 B ( A_2^2 +
A_1^2 \sin^2 2 \theta) ]/[ 24 (A_2 + A_1 \sin^2 2 \theta)^2 ] $.
\noindent The term $U_{o}$ is purely orientation-dependent without
a length scale, and is minimized at $\theta=$45$^o$ and 135$^o$,
as obtained in Ref.~\cite{Shenoy99}. The difference between
continuum and our discrete theories lies in the $k^2$ term:
continuum theory commonly assumes isotropic gradients in the order
parameter, i.e., $(\vec{\nabla} e_3)^2$~\cite{Barsch84}, whereas
$U_2(\theta)$ is anisotropic. The two origins of the anisotropy
are: (a) the compatibility relation, Eq.~(1), which has higher
powers in $k$ than Eq.~(3) due to discreteness, and (b) the
presence of shuffle mode energy. The latter can be written as
gradients of strains, but with corrections to the phenomenological
isotropic term, $(\vec{\nabla} e_3)^2$, commonly used in LG
theory. As $U_2(\theta)$ is minimized for $\theta=0^o$ and $90^o$,
it competes with $U_0(\theta)$  which prefers $\theta=45^o$ and
$135^o$. Thus, the domain wall direction depends on the length
scale with a critical length scale $\lambda_c \sim \sqrt{B/A_1}$.
If $\lambda_c \le 1$, i.e., less than the interatomic spacing, the
domain wall has direction 45$^o$ or 135$^o$ down to atomic scales.
If $\lambda_c>1$, then for length scales smaller (larger) than
$\lambda_c$, the domain wall direction is 0$^o$ or 90$^o$ (45$^o$
or 135$^o$) and the domain wall has multiscale attributes.

\begin{figure}
\leavevmode
\epsfxsize5.5cm\epsfbox{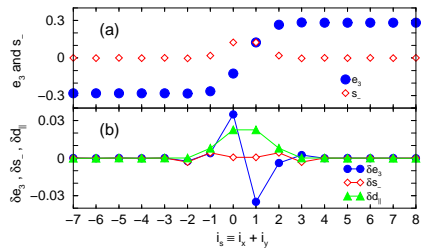}
\caption{\label{fig:wall.1d} (Color)
Atomic scale 135$^o$ domain wall profile for critical length scale,
$\lambda_c \le 1$
along the direction perpendicular to the domain wall:
(a) strain $e_{3}$ and shuffle $s_-$,
(b) differences in $e_{3}$
($\delta e_{3}$=$e_{3,\mathrm{atomic}}-e_{3,\mathrm{continuum}}$),
$s_-$ ($\delta s_{-}$) and
displacement parallel
to the domain wall direction ($\delta d_{||}$) between the results
from continuum theory
for ${\vec k} \sim 0$ and our model that
includes discreteness.
Parameter values are $A_1=5$, $A_2=4$, $A'_3=4$, $B=5$, and $F_3=50$.
}
\end{figure}
\begin{figure}
\centerline{\psfig{figure=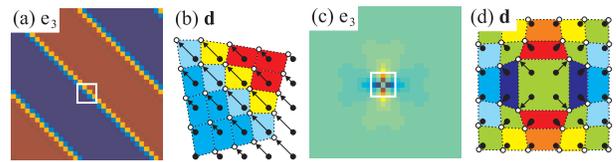,width=8.0cm,angle=0}} \caption{
Strain-$e_{3}$ mode for a periodic twinned microstructure (a) and
a single defect (c) together with their corresponding displacement
configurations [(b) and (d)] within the highlighted window.
$N_{L}=32\times 32$.} \label{FIG:Lattice}
\end{figure}

We  examine first the case $\lambda_c \le 1$ that would apply to
materials with relatively large bulk modulus $A_1$ (`hard' materials)
for fixed $B$.
Here $k_x=\pm k_y$ and $U(\vec{k})=B(1-\cos k_x)/(1+\cos k_x)$.
We illustrate the domain wall solution  with 135$^o$
domain wall direction.
The only non-zero
distortion modes
are $e_3$ and
$s_-$ ($s_+$ for a 45$^o$ domain wall).
The  strain $e_3$ reverses sign at the domain
wall, the intracell mode $s_-$ is confined within the domain wall,
and the atomic displacements are parallel to the domain wall direction.
The numerical
solution
 for $e_3$ and $s_-$ along a line perpendicular to the wall is shown
in Fig.~\ref{fig:wall.1d}(a), for which $\lambda_c \sim 1$. The
$e_{3}$-field and the corresponding displacement field near the
center of the domain wall are shown in Fig.~\ref{FIG:Lattice}(a)
and \ref{FIG:Lattice}(b), in which the red and blue colors  show
regions with $e_3$ positive and negative, respectively. Both
figures show that the center of the domain wall is located at
bonds rather than sites to avoid the higher energy state of
$e_3$=0 and large $s_-$. In Fig.~3(b) we  compare our results with
continuum theory, which predicts $e_3=e_3^{\text{max}} \tanh
(i_s/\xi )$~\cite{Barsch84} and $s_- = \partial e_3 /2 \partial
i_s $ from Eq.~(4), where $i_s=i_x+i_y$. The differences in the
interface region, shown in Fig.~\ref{fig:wall.1d}(b), are of the
order of $10\%$ of $e_3^{\text{max}}=\sqrt{A_3'/F_3}$.

\begin{figure}
\leavevmode \epsfxsize8.5cm\epsfbox{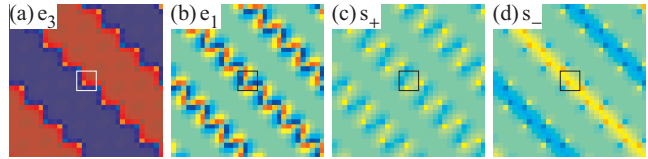}
\caption{\label{fig:wall.2d.step} (Color) Atomic scale domain wall
solution for materials with $\lambda_c>1$. Parameter values are
$A_1$=1, $A_2=4$, $A'_3=4$, $B=5$, and $F_3=50$. Strain $e_2$ is
zero. Dark red corresponds to 0.28, dark blue to -0.28, and green
to $\sim 0$. }
\end{figure}

The domain wall solution for  $\lambda_c>1$, typical for small bulk modulus
$A_1$ or `soft' materials,
is shown in Fig.~\ref{fig:wall.2d.step}
for which $\lambda_c \sim \sqrt{5}$.
The $e_3$ field in Fig.~\ref{fig:wall.2d.step}(a) shows that on length scales
of the size of the system (larger than $\lambda_c$),
the diagonal orientation is still preferred.
However, this diagonal domain wall consists of a
`staircase' of $0^o$ and $90^o$ domain walls of length scale $\lambda_c$.
More details on induced $e_1$, $s_+$, and $s_-$ fields
around the `staircase' wall in
Figs.~\ref{fig:wall.2d.step}(b)-\ref{fig:wall.2d.step}(d),
their implication for functionality of the domain walls,
and the displacement pattern are discussed in Ref.~\cite{Ahn02}.

A similar approach is used to find elastic texture around
structural defects. We consider impurity atoms at the centers of
the square of four atoms, which couples to the $e_1$ mode
distortion of the four nearest neighbor atoms. The corresponding
energy expression is $E_{sq,imp}=E_{sq.lat}+E_{imp}$, where
$E_{sq,lat}$ is Eq.~(\ref{eq:E.sq}) and $E_{imp}$ is
\begin{equation}
E_{imp}=\sum_{\vec{i}} C_1 e_1(\vec{i}) h_1(\vec{i}).
\end{equation}
Here, $h_1(\vec{i})$ is 1 if there is a defect at the site at
$\vec{i}+(1/2,1/2)$, and zero otherwise. $C_1$ represents the
strength of the coupling. $E_{sq,imp}$ is minimized about $e_1$,
$e_2$, $e_3$, $s_+$, and $s_-$ with constraints among them for
given $h_1$, which gives the relations between the relaxed strain
fields and the $h_1$ field. Explicit expressions of these
relations will be presented elsewhere. As a simple case, we show
the elastic texture around a {\it single} defect in
Figs.~\ref{FIG:Lattice}(c) and \ref{FIG:Lattice}(d).

To illustrate the influence of lattice deformation on  electronic
properties, we couple the twin boundary and defect solutions
obtained above with the electronic degree of freedoms in a model
of superconductors. The electronic model Hamiltonian is defined on
a square lattice:
\begin{eqnarray}
\mathcal{H}&=&-\sum_{ij,\sigma} \tilde{t}_{ij}
c_{i\sigma}^{\dagger}c_{j\sigma} +\sum_{i,\sigma}
(\epsilon_{i}-\mu) c_{i\sigma}^{\dagger}c_{i\sigma} \nonumber \\
&&+\sum_{ij} (\Delta_{ij} c_{i\uparrow}^{\dagger}
c_{j\downarrow}^{\dagger} +\Delta_{ij}^{*} c_{j\downarrow} c_{
i\uparrow} ) \;. \label{EQ:MFA}
\end{eqnarray}
Here  $c_{i\sigma}$ annihilates an electron of spin $\sigma$ on
site $i$. The quantities $\epsilon_{i}$ and $\mu$ are the on-site
impurity potential (if any) and the chemical potential,
respectively. The hopping integral  $\tilde{t}_{ij}$ is modified
by the lattice distortion. The electron-lattice coupling is
approximated by $t_{ij}=t_{ij}^{0}[1-\alpha \epsilon_{ij}]$, where
$t_{ij}^{0}$ is the bare hopping integral, $\epsilon_{ij}$ is the
lattice-distortion variable, and $\alpha$ is the coupling
constant. In our nearest neighbor realization, the bare hopping
integral $t_{ij}^{0}$ is $t$ for nearest neighbor sites and zero
otherwise. Specifically, we take the form of the lattice
distortion to be: $\epsilon_{ij}=[\vert
(\mathbf{R}_{j}+\mathbf{d}_{j})-
(\mathbf{R}_{i}+\mathbf{d}_{i})\vert/\vert
\mathbf{R}_{j}-\mathbf{R}_{i}\vert -1]$, where
$\{\mathbf{R}_{i}\}$ are the undistorted lattice coordinates and
$\{\mathbf{d}_{i}\}$ the lattice displacement vectors with respect
to $\{\mathbf{R}_{i}\}$. We assume an effective superconducting
gap function given by $\Delta_{ij}=\frac{U_{ij}}{2}\langle
c_{i\uparrow} c_{j\downarrow}
-c_{i\downarrow}c_{j\uparrow}\rangle$, where $U_{ij}=U\delta_{ij}$
(i.e., attractive Hubbard-$U$ model) for $s$-wave
superconductivity and $U_{ij}=V\delta_{i+\gamma,j}$ (with $\gamma$
specifying the nearest neighbors to the $i$-th site) for $d$-wave
superconductivity.
By performing a Bogoliubov-Valatin transformation, we may
diagonalize Eq.~(\ref{EQ:MFA}) by solving the Bogoliubov-de Gennes
(BdG) equation~\cite{deGennes89}:
\begin{equation}
\sum_{j} \left(
\begin{array}{cc}
{\cal H}_{ij} & \Delta_{ij}  \\
\Delta_{ij}^{*} & -{\cal H}_{ij}^{*}
\end{array}
\right) \left(
\begin{array}{c}
u_{j}^{n} \\ v_{j}^{n}
\end{array}
\right) =E_{n} \left(
\begin{array}{c}
u_{i}^{n} \\ v_{i}^{n}
\end{array}
\right)  \;, \label{EQ:BdG}
\end{equation}
subject to the self-consistency conditions for the superconducting
(SC) order parameter (OP):
\begin{equation}
\Delta_{ij}=\frac{U_{ij}}{4}\sum_{n} (u_{i}^{n}v_{j}^{n*}
+v_{i}^{n*}u_{j}^{n} ) \tanh \left(
\frac{E_{n}}{2k_{B}T}\right)\;. \label{EQ:Self-consistency}
\end{equation}
Here the single particle Hamiltonian reads ${\cal
H}_{ij}=-\tilde{t}_{ij} + (\epsilon_{i}-\mu)\delta_{ij}$.
We numerically solve the BdG equations self-consistently.
Below, we report results for two types of local lattice
distortions at zero temperature--- a superlattice formed by twin
boundaries and a single defect~\cite{Zhu03}. We measure the length
and energy in units of $a_0$ (the undistorted lattice constant)
and $t$. The chemical potential  $\mu=0$ and no extrinsic impurity
scattering is introduced ($\epsilon_{i}=0$). The pairing
interaction for both the $s$-wave ($U$) and $d$-wave ($V$)
superconductors is taken to be 3. The typical system size is
$N_{L}=32\times 32$ with periodic boundary conditions. When the
local quasiparticle density of states (LDOS) is computed, we
implement a much larger system using the above small system as a
supercell.

\begin{figure}
\centerline{\psfig{figure=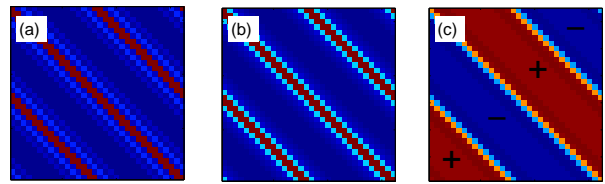,width=8.0cm,angle=0}}
\caption{Spatial variation of the SC OP for periodic twin
boundaries displayed in Fig.~\ref{FIG:Lattice}(a)--- (a) The
$s$-wave OP in an $s$-wave superconductor, and (b) the $d$-wave
and (c) extended $s$-wave components of the OP in a $d$-wave
superconductor. The electron-lattice coupling constant
$\alpha=3$.}
 \label{FIG:OP-TB}
\end{figure}

In Fig.~\ref{FIG:OP-TB}, we show the spatial variation of the SC
OP induced by the deformation of Fig.~\ref{FIG:Lattice}(a) in both
$s$ and $d$-wave superconductors. In both cases, the OP is lowered
within the domain and is elevated at the domain wall
(Fig.~\ref{FIG:OP-TB}(a-b)).
The magnitude of the OP is depressed in comparison to an
undistorted square lattice since the lattice deformation changes
the band structure, leading to a reduction in  normal density of
states at the Fermi energy. Even at the domain wall, where the
strain induced deformation is weakest, the amplitude of the
enhanced OP is smaller than its value in an undistorted
 square lattice. This is due to the confinement from the
two neighboring domains.
In a twinned
domain of a $d$-wave SC, a subdominant extended $s$-wave component
is generated in a real combination $d\pm s$. Because the
symmetries of two twinned domains are reflected into each other
with respect to the twin boundary, the relative phase between the
$d$- and $s$-wave components switches by $\pi$  when a twin
boundary is crossed (Fig.~\ref{FIG:OP-TB}(c)).

\begin{figure}
\centerline{\psfig{figure=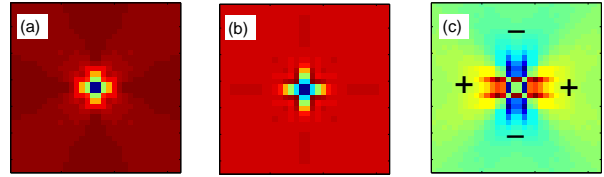,width=8.0cm,angle=0}}
\caption{Spatial variation of the SC OP for a single defect
displayed in Fig.~\ref{FIG:Lattice}(c)--- (a) The $s$-wave OP in
an $s$-wave superconductor, and (b) the $d$-wave and (c) extended
$s$-wave components of the OP in a $d$-wave superconductor. The
electron-lattice coupling constant $\alpha=3$. }
 \label{FIG:OP-DF3}
\end{figure}

We show in Fig.~\ref{FIG:OP-DF3} the spatial variation of the
superconducting OP around the single defect
(Fig.~\ref{FIG:Lattice}(c)) in both the $s$-wave and $d$-wave
superconductor cases. The OP is depressed at the center of the
defect, and reaches its defect-free bulk value at the scale of the
superconducting coherence length. Notice that for a
lattice-deformation defect, which affects the local electron
hopping integral, the OP has a minimum at four sites surrounding
the defect center. It is different from the case of an externally
substituted unitary impurity, where the minimum OP is located only
at the impurity site itself~\cite{Zhu00}. The range of influence
of such a defect can be very large depending on the strength of
electron-lattice coupling--- {\em the elasticity propagates the
electronic response}. The $d$-wave energy gap has a sign change at
the nodal directions of the essentially cylindrical Fermi surface,
but the $d$-wave OP does not exhibit such a sign change in real
space. With the defect, an extended $s$-wave component of the OP
is induced when the dominant $d$-wave component is depressed at
the defect. Strikingly, the induced $s$-wave component has a sign
change across the diagonals of the square lattice, i.e.,
$\text{sgn}[\cos(2\theta)]$, where $\theta$ is the azimuthal angle
with respect to the crystalline $x$ axis. This is a direct
manifestation of the $d$-wave pairing symmetry in real space.

\begin{figure}
\centerline{\psfig{figure=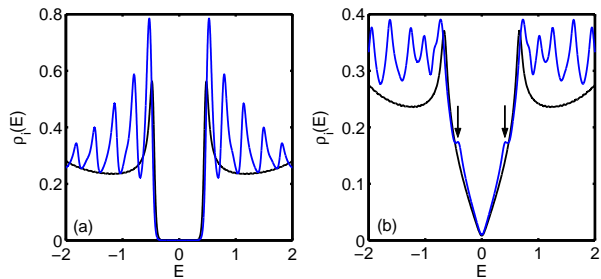,width=8.0cm,angle=0}}
\caption{The LDOS at a twin boundary in $s$-wave (a) and $d$-wave
(b) superconductors. Also shown are the LDOS (black lines) for a
uniform domain. The electron-lattice coupling constant
$\alpha=3$.} \label{FIG:LDOS-TB}
\end{figure}

Once the self-consistency for the order parameter is obtained, we
calculate the LDOS:
\begin{equation} \rho_{i}(E)= -\sum_{n}[ \vert
u_{i}^{n} \vert^{2} f^{\prime}(E-E_{n}) + \vert v_{i}^{n}
\vert^{2} f^{\prime}(E+E_{n})]\;,
\end{equation}
where $f^{\prime}(E)$ is the derivative of the Fermi distribution
function with respect to the energy. The LDOS determines the
differential tunneling conductance, measurable by STM
experiments~\cite{Pan00}.
Figure~\ref{FIG:LDOS-TB} shows the LDOS at a domain wall for both
types of superconductors, where the modulation of the OP forms a
superlattice, with maximum at the domain wall playing the role of
an off-diagonal potential barrier ($\Delta_{ij}$ in
Eq.~(\ref{EQ:BdG})). For an $s$-wave superconductor, the
quasiparticles are gapped away with their energy below the minimum
SC OP. Outside the minimum of the pair potential, energy bands are
formed by the quasiparticle scattering off the off-diagonal energy
barriers at the domain walls.
Interestingly, the bottom of the oscillation pattern follows the
LDOS (black line) of a system formed by a uniform rectangular
domain. Similar oscillations are obtained for the $d$-wave
superconductor. However, the bottom of the oscillations do not
follow the single domain DOS (black line). In addition, weak
subgap peaks (labeled by arrows in Fig.~\ref{FIG:LDOS-TB}(b))
appear symmetrically in the LDOS on the domain wall but are absent
in the single-domain LDOS. We speculate that  these resonant
states  are due to the gradient of the $s$-wave gap component
induced inside the domain.

\begin{figure}
\centerline{\psfig{figure=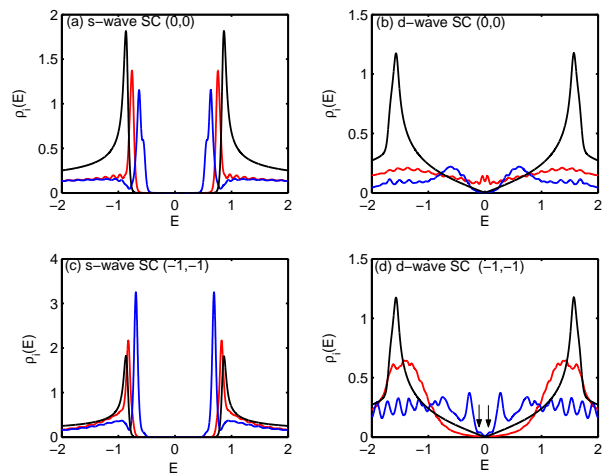,width=8.0cm,angle=0}}
\caption{The LDOS near the center of a defect in $s$-wave (left
column) and $d$-wave (right column) superconductors. The distance
of the measured point away from the defect is labeled by its
coordinate. The electron-lattice coupling constants are $\alpha=3$
(red lines) and $10$ (blue lines). Also shown is the defect-free
LDOS (black lines). } \label{FIG:LDOS-DF3}
\end{figure}

In Fig.~\ref{FIG:LDOS-DF3}, we show the calculated LDOS near the
center of a single defect. The depression of the SC OP at the
defect makes a quantum-well-like profile of the energy gap. The
size and depth of the well is determined by the
electron-lattice coupling constant. 
In the $s$-wave superconductor, the well is shallow and small for
weak coupling, which cannot trap low-lying quasiparticle bound
states; for strong coupling constants, the well is deep and large
so that subgap quasiparticle bound states are induced (the red and
blue lines of Figs.~\ref{FIG:LDOS-DF3}(a) and (c)). The energy of
these low-lying states must be inbetween the bottom and edge of
the well. Therefore, it is notable that the energy of these subgap
states is shifted toward the Fermi surface as the electron-lattice
coupling is increased (the blue line in
Figs.~\ref{FIG:LDOS-DF3}(a) and (c)).
The electronic structure at the defect in a $d$-wave
superconductor becomes even richer: For $\alpha=3$ (weak coupling
as compared to the band width of the uniform square lattice), the
lattice distortion plays the role of a weak defect for the
quasiparticle scattering.
In this case, a resonant peak with a dip exactly at the Fermi
energy is seen (the red line in Fig.~\ref{FIG:LDOS-DF3}(b)). The
overall peak comes from the scattering of quasiparticles off the
single-particle off-diagonal potential (i.e., local change of the
hopping integral as a response to the lattice deformation). This
lattice-deformation induced resonance state also exhibits Friedel
oscillations. Typically, the peak structure appears in the LDOS at
(0,0) (We label the four sites surrounding the defect center by
(0,0), (1,0), (1,1), (0,1)) and (-2,-2).
For $\alpha=10$ (strong coupling),
the `resonant' peaks are pushed to higher energies ($\simeq \pm
0.3$) (the blue line of Fig.~\ref{FIG:LDOS-DF3}(d)). Furthermore,
small shoulders appear close to the Fermi energy (the blue lines
of Figs.~\ref{FIG:LDOS-DF3}(b) and (d)), which are precursors of
new Andreev resonance states. We have also computed the LDOS
without imposing self-consistency on the OP and found that the
double-peak structure is V-shaped with no existence of the
shoulders. Therefore,  the new Andreev resonance states originate
from the confinement of the induced $s$-wave OP.
All these features are unique to an elastic defect in a
$d$-wave superconductor with short coherence length.

In summary, we have reported an approach to ``atomic-scale
elasticity'', which uses symmetry modes of elementary objects of
atoms as distortion variables. A gradient expansion for the energy
with {\it anisotropic} coefficients has been obtained, with
corrections to the usual phenomenological isotropic gradient terms
used in LG theory. As an illustration, we obtained domain wall
(twin boundary) solutions and elastic texture around a defect in
terms of strain {\it and} intracell modes, and showed how the
domain wall solutions  differ from the continuum elastic soliton
solution~\cite{Barsch84}. Using the atomic scale profiles of
elastic texture, we studied the effects of elastic lattice
deformation on the nanoscale electronic structure in
superconductors within a BdG approach. We showed that the SC OP is
depressed in the regions where the lattice deformation exists. The
calculated LDOS suggests that the electronic structure is strongly
modulated in response to the lattice deformation. In particular,
it is possible to trap low-lying quasiparticle states around the
defects. Images of these states will manifest the underlying
long-range anisotropic elastic
 lattice deformation. These predictions can be directly
tested by STM experiments in new superconducting materials. Our
approach is readily extended to other elastic textures, SC
symmetries, and lattices, as well as coupling to other electronic
models (for charge-transfer, charge-density-wave,
spin-density-wave, Jahn-Teller, etc.).

We thank A. J. Millis and S. R. Shenoy for insightful discussions.
This work was supported by the US Department of Energy.

\end{document}